\documentclass[12pt]{article}
\usepackage{axodraw,epsfig}
\usepackage{graphics}

\newcommand{\bfm}[1]{\mbox{\boldmath$#1$}}

\catcode`@=11
\newcount\@tempcntc
\def\@citex[#1]#2{\if@filesw\immediate\write\@auxout{\string\citation{#2}}\fi
  \@tempcnta\z@\@tempcntb\m@ne\def\@citea{}\@cite{\@for\@citeb:=#2\do
    {\@ifundefined
       {b@\@citeb}{\@citeo\@tempcntb\m@ne\@citea\def\@citea{,}{\bf ?}\@warning
       {Citation `\@citeb' on page \thepage \space undefined}}%
    {\setbox\z@\hbox{\global\@tempcntc0\csname b@\@citeb\endcsname\relax}%
     \ifnum\@tempcntc=\z@ \@citeo\@tempcntb\m@ne
       \@citea\def\@citea{,}\hbox{\csname b@\@citeb\endcsname}%
     \else
      \advance\@tempcntb\@ne
      \ifnum\@tempcntb=\@tempcntc
      \else\advance\@tempcntb\m@ne\@citeo
      \@tempcnta\@tempcntc\@tempcntb\@tempcntc\fi\fi}}\@citeo}{#1}}
\def\@citeo{\ifnum\@tempcnta>\@tempcntb\else\@citea\def\@citea{,}%
  \ifnum\@tempcnta=\@tempcntb\the\@tempcnta\else
   {\advance\@tempcnta\@ne\ifnum\@tempcnta=\@tempcntb \else \def\@citea{--}\fi
    \advance\@tempcnta\m@ne\the\@tempcnta\@citea\the\@tempcntb}\fi\fi}
\catcode`@=12

\begin{document}

\begin{titlepage}

    \begin{flushright}
      \normalsize PITHA~07/03\\
      \normalsize TTP/07-11\\
      \normalsize SFB/CPP-07-22\\
      18 June 2007  
    \end{flushright}

\vskip1.5cm
\begin{center}
\Large\bf\boldmath
 Ultrasoft contribution to quarkonium production and annihilation
\unboldmath
\end{center}

\vspace*{0.8cm}
\begin{center}

{\sc M. Beneke}$^{a}$,
{\sc Y. Kiyo}$^{b}$, 
{\sc A.A. Penin}$^{b,c,d}$\\[5mm]
  {\small\it $^a$ Institut f{\"u}r Theoretische Physik~E,
    RWTH Aachen,}\\
  {\small\it D--52056 Aachen, Germany}\\[0.1cm]
  {\small\it $^b$ Institut f{\"u}r Theoretische Teilchenphysik,
    Universit{\"a}t Karlsruhe,}\\
  {\small\it D--76128 Karlsruhe, Germany}\\[0.1cm]
  {\small\it $^c$ Department of Physics,
  University Of Alberta}\\
  {\small \it Edmonton, AB T6G 2J1, Canada}\\[0.15mm]
  {\small\it $^d$ Institute for Nuclear Research,
    Russian Academy of Sciences,}\\
  {\small\it 119899 Moscow, Russia}
        
\end{center}

\vspace*{0.8cm}
\begin{abstract}
  \noindent
  We compute the third-order correction to electromagnetic 
  $S$-wave quarkonium production and annihilation rates 
  due to the emission and
  absorption of an ultrasoft gluon. Our result completes the
  analysis of the non-relativistic quarkonium bound-state dynamics in the
  next-to-next-to-next-to-leading  order. The impact of the
  ultrasoft correction on the $\Upsilon(1S)$ leptonic width and the top
  quark-antiquark threshold production cross section is 
  estimated.

\vspace*{0.8cm}

\end{abstract}

\vfil
\end{titlepage}

\newpage

\section{Introduction}

The theoretical study of non-relativistic heavy quark-antiquark
systems is among the earliest applications of perturbative quantum
chromodynamics (QCD) \cite{AppPol}. Perturbation theory applies 
to the bound-state dynamics of bottomonium, 
at least within the sum rule approach
\cite{NSVZ}, and top-antitop systems \cite{FadKho}, since 
non-perturbative effects are under control \cite{Vol,Leu}.
This makes heavy quark-antiquark systems well suited to
determine fundamental parameters of QCD, the strong coupling
constant $\alpha_s$ and the heavy-quark masses $m$.

The binding energy of a quarkonium state and the value of its
wave function at the origin -- field-theoretically, the residues of 
two-point functions of local currents -- 
are of primary phenomenological interest.  The
former determines the mass of the bound state, while the
latter controls its production and annihilation rates.  The 
quarkonium ground-state energy has been computed through ${\cal
  O}(m \alpha_s^5)$ including the third-order correction to the Coulomb
approximation~\cite{KPSS1,PenSte}. This result has been extended
to the excited $S$-wave states \cite{PSS,BKS1}.  For the wave function
at the origin a complete result is only available 
including the second-order correction~\cite{MelYel1,PenPiv1,BSS2}. 
This correction is large even for 
top quarks, and for a reliable perturbative
prediction the third-order approximation seems to be needed.  
This amounts to a difficult calculation, which can be broken 
into several well-defined pieces, some of which are already 
available, such as the double-logarithmically enhanced ${\cal
  O}(\alpha_s^3\ln^2\alpha_s)$ terms \cite{KniPen2,ManSte} and the
single-logarithmic ${\cal O}(\alpha_s^3\ln\alpha_s)$ terms 
\cite{KPSS2,Hoa2}.  The calculation of the most difficult 
non-logarithmic term has been started in~\cite{PSS,BKS1}, where the 
contribution to the wave function at the origin from the loop 
corrections to the colour-Coulomb potential have been
evaluated. In this and the companion paper~\cite{BKS2} the 
remaining contributions from the non-Coulomb potentials and 
due to the emission and absorption of an ultrasoft gluon 
by the quarkonium bound state are presented. The ultrasoft 
correction discussed below is of special interest, because
it constitutes a qualitatively new effect, which shows up for the 
first time in the third order. No other such effects are expected in higher
orders of the perturbative expansion. The complete third-order 
correction to the wave function at the origin can now be 
expressed in terms of a few yet unknown matching coefficients, 
which can be obtained by standard fixed-order loop calculations.

\section{Ultrasoft correction to the 
wave function} 

\subsection{Definitions}
\label{sec1}

In non-relativistic bound states the quark velocity $v$ is a
small parameter. An expansion in $v$ may be performed directly in
the QCD Lagrangian by using the framework of effective field theory
\cite{CasLep,BBL,PinSot1}, or diagrammatically with  
the threshold expansion~\cite{BenSmi}. The relevant momentum 
regions are the hard region (energy $k^0$ and momentum $\bfm k$ 
of order $m$), the soft region ($k^0,{\bfm k} \sim m v$), 
the potential region ($k^0 \sim mv^2$, ${\bfm k}\sim m v $), 
and the ultrasoft region ($k^0,{\bfm k} \sim m v^2$).
Integrating out the hard modes amounts to matching onto 
non-relativistic QCD (NRQCD) \cite{BBL}.  If
one also integrates out the soft modes and potential gluons, one
obtains the effective theory called potential NRQCD (PNRQCD), which
contains potential heavy quarks and ultrasoft gluons as dynamical
fields~\cite{PinSot1} (see also~\cite{B98}). 
In this theory the leading colour-Coulomb 
potential is part of the unperturbed Lagrangian, so that 
the propagation of a colour-singlet 
quark-antiquark pair is described by the Green 
function of the Schr{\"o}dinger equation
\begin{equation}
\left(H_0-E\right)G_C^{(s)}({\bfm r},{\bfm r}';E) =\delta^{(3)}({\bfm
r}-{\bfm r}')\,,
\label{schroed}
\end{equation}
with \begin{eqnarray}
H_0&=& -{\bfm \nabla_{\!(r)}^2\over m}-\frac{\alpha_s C_F}{r},
\label{hamilton}
\end{eqnarray}
$r=|{\bfm r}|$, $m$ the heavy-quark pole mass, 
and $C_F=(N_c^2-1)/(2N_c)$, $N_c=3$. 
The \mbox{PNRQCD} Lagrangian further contains interactions of quarks 
with the multipole-expanded ultrasoft gluon field and 
instantaneous, spatially non-local interactions (``potentials''), 
which can be treated as perturbations. This constitutes the basic
framework for the perturbative analysis of quarkonium bound-state 
properties. The colour-singlet Coulomb Green function
$G_C^{(s)}({\bfm r},{\bfm r}';E)$ has an infinite number 
of bound-state poles with energies $E_n^{(0)} = -m (\alpha_s C_F)^2/
(2 n)^2$ and contains the information about the corresponding 
wave functions. In the quark-antiquark Fock state sector the 
perturbations to the energy levels and wave functions can 
be taken into account by replacing $H_0$ by the PNRQCD 
Hamiltonian $H$ with the ultrasoft modes are excluded. 

In this paper, however, we are interested in the leading 
ultrasoft effect.  
To connect the concept of a non-relativistic wave function 
at the origin to a physical quantity, we 
consider the two-point function of the electromagnetic
heavy-quark current $j_\mu=\bar Q\gamma_\mu Q$ in full QCD, 
\begin{equation}
\left(q_\mu q_\nu-g_{\mu \nu}q^2\right)\Pi(q^2)=
i\int d^dx\,e^{iqx}\,\langle 0|Tj_{\mu}(x)j_{\nu}(0)|0\rangle \,,
\label{vacpol}
\end{equation}
whose poles are related to electromagnetic production
and annihilation rates of the corresponding 
quarkonium states. In PNRQCD $j_\mu$ is represented in
terms of operators constructed from the non-relativistic quark and
antiquark two-component Pauli spinor fields $\psi$ and $\chi$,
\begin{equation}
{\bfm j}=c_v\psi^\dagger{\bfm \sigma}\chi+{d_v\over6 m^2}
\psi^\dagger{\bfm \sigma}\mbox{\bfm D}^2\chi
+\ldots.
\label{currentnr}
\end{equation}
The matching coefficients 
$c_v(\mu)=1+c^{(1)}_v(\mu)\alpha_s(\mu)/(4\pi)+\ldots$ and
$d_v(\mu)=1+{\cal O}(\alpha_s)$ represent the
contributions from the hard modes with $\mu$ a factorization scale 
that is also implicit in the renormalization convention for 
the operators on the right-hand side. (We use dimensional regularization 
with $d=4-2\epsilon$ and the $\overline{\rm MS}$ scheme.) 
We also introduce the PNRQCD two-point function 
\begin{equation}
2 (d-1) N_c \,G(E) = 
i\int d^dx\,e^{iE x^0}\,\langle 0|T [\psi^\dagger\sigma^i\chi](x)
\,[\chi^\dagger\sigma^i\psi](0)|0\rangle \,,
\label{defG}
\end{equation}
where $E=\sqrt{q^2}-2m$. In leading order $G(E)$ coincides 
with (the correspondingly regularized) $G_C^{(s)}(0,0;E)$. Substituting the 
expansion~(\ref{currentnr}) into (\ref{vacpol}) and using an 
equation-of-motion relation for the insertion of the derivative current 
in~(\ref{currentnr}), we obtain 
\begin{equation}
\Pi(q^2)
={N_cc_v\over 2 m^2}\left[
c_v-{E\over m}\left(1+{d_v\over 3}\right)+\ldots\right] G(E)\,,
\label{vacpolnr}
\end{equation}
which is valid up to the third order. $G(E)$ has Coulomb bound-state 
poles at energies $E_n\approx E_n^{(0)}$ 
with spin and orbital angular momentum 
$S=1$ and $l=0$, respectively, following from the form of the 
current, $\psi^\dagger {\bfm \sigma}\chi$. Near the pole
\begin{equation}
G(0,0;E)~\raisebox{0pt}{$\longrightarrow$}
\raisebox{-5pt}{$\hspace*{-22pt}_{E\to E_n}$}~{|\psi_n(0)|^2\over
  E_n-E-i\varepsilon}
\, ,
\label{gfpole}
\end{equation}
which defines the ``wave function at the origin''.  Beginning from 
the second-order correction, these wave-functions are 
factorization-scale dependent, but this dependence cancels in the 
residues of the poles of $\Pi(q^2)$, which determine the observable 
electromagnetic production and annihilation rates.
 
The third-order (NNNLO) corrections to $|\psi_n(0)|^2$ originate 
(i) from single insertions of the third-order potentials in the 
PNRQCD Lagrangian and multiple insertions of first- and second-order 
potentials, where the order is determined by the combined 
suppression in $\alpha_s$ and $v$ relative to the leading-order 
Coulomb potential $\alpha_s/r \sim \alpha_s v$. This contribution 
has been computed in~\cite{PSS,BKS1,BKS2} and expressed in terms 
of the yet unknown three-loop correction to the Coulomb potential 
and four constants related to ${\cal O}(\epsilon)$ parts of 
loop corrections to other potentials; (ii) from the 
emission and absorption of an ultrasoft gluon (similar to the 
Lamb shift for energy levels), calculated below. 

\subsection{Calculation of the ultrasoft correction}
\label{sec22}

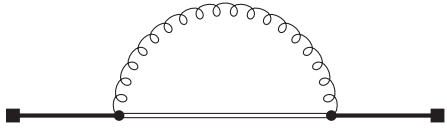
\begin{figure}[t]
\vspace*{-0.2cm}
\begin{center}
\begin{picture}(300,60)(0,0)
\SetScale{0.8}
\SetWidth{2} \Line(50,0)(100,0) \Line(200,0)(250,0) \SetWidth{0.5}
\Line(100,1)(200,1) \Line(100,-1)(200,-1)
\GlueArc(150,00)(50,0,180){3}{18.5} \Vertex(100,0){2.5}
\Vertex(200,0){2.5} 
\CBoxc(50,0)(6,6){Black}{Black}
\CBoxc(250,0)(6,6){Black}{Black}
\end{picture}
\end{center}
\caption{\label{fig1} \small The ultrasoft correction as a 
PNRQCD Feynman diagram. The bold and the double lines stand
for the singlet and octet Coulomb Green functions, respectively, the curly
line represents the ultrasoft-gluon propagator,
the  black circles  represent  the chromoelectric dipole interaction
$g_s{\bfm r}{\bfm E}$, and the squares correspond to the
non-relativistic currents.}
\end{figure}

The leading ultrasoft interactions in the PNRQCD Lagrangian are 
$g_s \psi^\dagger(x)(A^0(t,{\bfm 0}) - {\bfm x}{\bfm E}(t,{\bfm 0}))
\psi(x)$ together with a similar term for the antiquark field. 
The contribution from the $A^0$ coupling cancels (or can be gauged 
away), leaving the chromoelectric dipole interaction, which 
results in a NNNLO correction~\cite{KniPen1,BPSV,B99}. 
The PNRQCD diagram representing this correction is shown in 
Fig.~\ref{fig1}. The corresponding correction to $G(E)$ defined 
in (\ref{defG}) reads
\begin{eqnarray}
\delta^{us} G(E)&=& ig_s^2 C_F\int d^3{\bfm r}
d^3{\bfm r}'\int \frac{d^4{k}}{(2\pi)^4}\,
\Bigg[\frac{k_0^2\,{\bfm r}{\bfm r}'- ({\bfm r}{\bfm
      k})({\bfm r}'{\bfm k})}{k^2+i\varepsilon}
\nonumber\\
&&
\times\,G^{(s)}_C(0,{\bfm r};E)G^{(o)}_C({\bfm r},{\bfm r}';E-k_0)
G^{(s)}_C({\bfm r}',0;E)\Bigg]\,
\label{gfcorr}
\end{eqnarray}
with the understanding that one picks up only the pole 
at $k^0=|{\bfm k}|-i\epsilon$ in the gluon propagator. 
Here $G^{(o)}_C({\bfm r},{\bfm r}';E)$
denotes the colour-octet Coulomb Green function corresponding 
to the Hamiltonian (\ref{hamilton}) with the 
octet potential $V_C^{(o)}(r)=(C_A/2-C_F)\alpha_s/r$ ($C_A=N_c$). 
Only the $l=1$ partial wave of the octet Green function contributes to 
(\ref{gfcorr}), and since the octet potential is repulsive, 
$G^{(8)}_C({\bfm r},{\bfm r}';E)$ does not have bound-state poles.

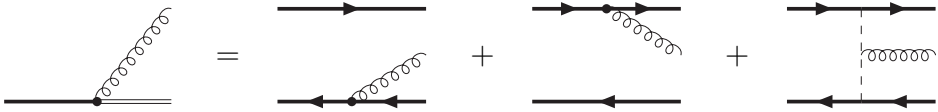
\begin{figure}[t]
\vspace*{-0.4cm}
\begin{center}
\hspace*{1.5cm}
\begin{picture}(100,60)(50,0)
\SetScale{0.7}
\SetWidth{2}
\Line(50,0)(100,0)
\SetWidth{0.5}
\Line(100,1)(140,1)
\Line(100,-1)(140,-1)
\Gluon(100,0)(140,50){3}{8.5}
\Vertex(100,0){2.5}
\end{picture}
\hspace*{-1.5cm}
\raisebox{5mm}{$=$}
\hspace*{0.5cm}
\begin{picture}(100,60)(50,0)
\SetScale{0.7}
\SetWidth{2}
\ArrowLine(100,0)(60,0)
\ArrowLine(60,50)(140,50)
\ArrowLine(140,0)(100,0)
\SetWidth{0.5}
\Gluon(100,0)(140,25){3}{6.5}
\Vertex(100,0){2.5}
\end{picture}
\hspace*{-1.5cm}
\raisebox{5mm}{$+$}
\hspace*{0.5cm}
\begin{picture}(100,60)(50,0)
\SetScale{0.7}
\SetWidth{2}
\ArrowLine(60,50)(100,50)
\ArrowLine(100,50)(140,50)
\ArrowLine(140,0)(60,0)
\SetWidth{0.5}
\Gluon(100,50)(140,25){3}{6.5}
\Vertex(100,50){2.5}
\end{picture}
\hspace*{-1.5cm}
\raisebox{5mm}{$+$}
\hspace*{0.5cm}
\begin{picture}(100,60)(50,0)
\SetScale{0.7}
\SetWidth{2}
\ArrowLine(100,0)(60,0)
\ArrowLine(60,50)(100,50)
\ArrowLine(100,50)(140,50)
\ArrowLine(140,0)(100,0)
\SetWidth{0.5}
\DashLine(100,0)(100,50){5}
\Gluon(100,25)(140,25){3}{6.5}
\end{picture}
\end{center}
\caption{\label{fig2} \small
  The NRQCD decomposition of the PNRQCD vertex.  The dashed line
  corresponds to a potential Coulomb gluon.  The black circles on the
  right-hand side correspond to the NRQCD vertices $g_s{\bfm p}{\bfm
    A}/m$. The bold arrows correspond to the potential quark and
  antiquark propagators with any number of the potential gluon  exchanges.}
\end{figure}

Expression (\ref{gfcorr}) cannot be used in practice, because the 
ultrasoft correction is divergent. Its definition requires specifying
a regulator and subtractions, which must be chosen to be consistent
with the calculation of the potential insertions; the potentials 
themselves; and the hard matching coefficients. In the following 
we provide the necessary definitions, deferring a detailed discussion 
of many interesting technical aspects of the calculation to a 
later publication that will provide results not only 
for $|\psi_n(0)|^2$ but also for the full correlation function $G(E)$. 
To apply dimensional regularization we transform (\ref{gfcorr}) 
to momentum space. We also find it convenient to re-express 
the PNRQCD vertex in terms of the NRQCD vertices from which 
it is derived by equation-of-motion relations, see~\cite{B99} 
and Fig.~\ref{fig2} for a graphical representation. The formulation of
non-relativistic effective theory in dimensional 
regularization~\cite{KPSS1,BSS2,PinSot2,CMY} is very convenient, 
because it allows one to combine bound-state calculations 
with loop calculations of matching coefficients, which are 
technically feasible only in dimensional regularization. Furthermore, 
when loop integrals are expanded in the sense   of the 
threshold expansion \cite{BenSmi}, the matching of contributions 
from different regions is automatic.

The integral over the three-momentum $\bfm k$ of the ultrasoft gluon is
ultraviolet (UV) divergent. The divergence is related to the
factorization of the ultrasoft scale from the other scales, 
and cancels when all pieces of the calculation are added. The
UV-divergent part of the ultrasoft integral has the form
of a single insertion of a third-order potential and of a one-loop
correction to the coefficient $d_v$ of the 
derivative current in (\ref{currentnr}). 
In fact, they cancel precisely infrared (IR) divergences in 
the calculation of these quantities \cite{KPSS1,LukSav}.
We therefore define the ultrasoft 
correction by adding counterterms that cancel these ultrasoft 
subdivergences, 
\begin{eqnarray}
\delta^{us} G(E)&=& 
\Big[\tilde\mu^{2\epsilon}\Big]^2\int \frac{d^{d-1}{\bfm \ell}}{(2\pi)^{d-1}}
\frac{d^{d-1}{\bfm \ell ^\prime}}{(2\pi)^{d-1}}
\Bigg\{(-\delta d_v^{\rm div})\,\frac{\bfm{\ell^2}+\bfm{\ell^{\prime\,2}}}
{6 m^2}\,\tilde G^{(s)}_C({\bfm \ell},{\bfm \ell^\prime};E)
\nonumber\\
&& +\,\Big[\tilde\mu^{2\epsilon}\Big]^2
\int \frac{d^{d-1}{\bfm p}}{(2\pi)^{d-1}}
\frac{d^{d-1}{\bfm p^\prime}}{(2\pi)^{d-1}}
\,\tilde G^{(s)}_C({\bfm \ell},{\bfm p};E)
\Big[\delta U +\delta\tilde V_{c.t.}\Big]
\,\tilde G^{(s)}_C({\bfm p^\prime},{\bfm \ell^\prime};E)
\Bigg\},
\label{defUS}
\end{eqnarray}
where $\delta U$ follows from (\ref{gfcorr}), and the counterterms 
read (${\bfm q} = {\bfm p}-\bfm{p^\prime}$)
\begin{eqnarray}
\delta \tilde V_{c.t.} &=&{\alpha_s C_F\over 6\epsilon}
\Bigg[C_A^3{\alpha_s^3\over {\bfm
    q}^2}+4 (C_A^2+2C_AC_F)
    {\pi \alpha_s^2\over m|{\bfm q}|}
\nonumber\\
&&
  +16\left(C_F-\frac{C_A}{2}\right){\alpha_s\over
  m^2} +16C_A{\alpha_s\over
  m^2} {{\bfm p}^2+{\bfm p}'^2\over{2 \bfm q}^2}
\Bigg]\,,
\\
\delta d_v^{\rm div} &=&-\frac{\alpha_s}{4\pi}\,
\frac{16 C_F}{\epsilon}.
\end{eqnarray}
The counterterms added here are subtracted from the other 
parts of the calculation \cite{BKS2}. With $\tilde\mu^2 = 
e^{\gamma_E}\,\mu^2/(4\pi)$ subtracting poles in $\epsilon$ 
corresponds to the $\overline{\rm MS}$ subtraction.

\begin{figure}[t]
\vspace*{-0.2cm}
\begin{center}
\hspace*{1.2cm}
\begin{picture}(90,60)(70,0)
\SetScale{0.7}
\ArrowLine(100,0)(70,25)
\ArrowLine(70,25)(100,50)
\ArrowLine(130,0)(100,0)
\ArrowLine(100,50)(130,50)
\SetWidth{2}
\ArrowLine(160,0)(130,0)
\ArrowLine(130,50)(160,50)
\SetWidth{0.5}
\DashLine(100,0)(100,50){5}
\DashLine(130,0)(130,50){5}
\Gluon(100,25)(130,25){3}{5.5}
\CBoxc(70,25)(6,6){Black}{Black}
\end{picture}
\hspace*{-1.5cm}
\raisebox{5mm}{$=$}
\hspace*{1cm}
\begin{picture}(100,60)(65,0)
\SetScale{0.7}
\ArrowLine(100,0)(70,25) \ArrowLine(70,25)(100,50)
\ArrowLine(135,0)(100,0) \ArrowLine(100,50)(135,50) \SetWidth{2}
\ArrowLine(170,0)(140,25) \ArrowLine(140,25)(170,50) \SetWidth{0.5}
\DashLine(100,0)(100,50){5} \DashLine(130,0)(130,50){5}
\Gluon(100,25)(130,25){3}{5.5} 
\CBoxc(70,25)(6,6){Black}{Black} 
\Text(102,46)[cc]{\footnotesize $\bfm{p}=0$} 
\end{picture}
\hspace*{-1.5cm}
\raisebox{5mm}{$+$}
\hspace*{1cm}
\begin{picture}(90,60)(65,0)
\SetScale{0.7}
\ArrowLine(100,0)(70,25)
\ArrowLine(70,25)(100,50)
\ArrowLine(130,0)(100,0)
\ArrowLine(100,50)(130,50)
\SetWidth{2}
\ArrowLine(160,0)(130,0)
\ArrowLine(130,50)(160,50)
\SetWidth{0.5}
\DashLine(100,0)(100,50){5}
\DashLine(130,0)(130,50){5}
\Gluon(100,25)(130,25){3}{5.5}
\CBoxc(70,25)(6,6){Black}{Black}
\SetWidth{0.9}
\CArc(123,25)(68,151,209)
\SetWidth{0.5}
\end{picture}
\hspace*{-1.5cm}
\raisebox{5mm}{$-$}
\hspace*{1cm}
\begin{picture}(80,60)(65,0)
\SetScale{0.7}
\ArrowLine(100,0)(70,25) \ArrowLine(70,25)(100,50)
\ArrowLine(135,0)(100,0) \ArrowLine(100,50)(135,50) \SetWidth{2}
\ArrowLine(170,0)(140,25) \ArrowLine(140,25)(170,50) \SetWidth{0.5}
\DashLine(100,0)(100,50){5} \DashLine(130,0)(130,50){5}
\Gluon(100,25)(130,25){3}{5.5} 
\CBoxc(70,25)(6,6){Black}{Black}
\Text(102,46)[cc]{\footnotesize $\bfm{p}=0$} 
\SetWidth{0.9} 
\CArc(117,25)(68,-29,29) \SetWidth{0.5}
\end{picture}
\end{center}
\caption{\label{fig3} \small
  Example of a three-loop vertex correction with an ultrasoft
  exchange.  The right-hand side illustrates the subtraction of the UV
  divergent part.  The thin arrows correspond to free potential
  quark and antiquark propagators.}
\end{figure}
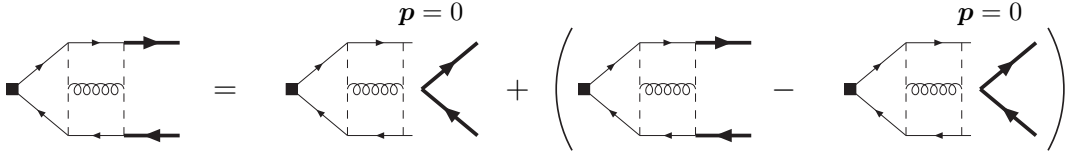

After adding this counterterm the potential loops and Coulomb Green
functions can be evaluated in three dimensions unless the 
potential loop integrations are divergent. Such divergences 
occur in potential vertex subgraphs up to three loops, and 
they correspond to 
IR divergences in the three-loop correction to $c_v$. 
(There are also over-all divergences in $G(E)$, but they 
are irrelevant to the calculation of the bound-state pole and 
residue.) We separate this vertex subdivergence by adding and 
subtracting the three-loop vertex
subdiagram at zero external momentum ${\bfm p}$, as shown
in Fig.~\ref{fig3}. The vertex UV divergence is logarithmic, 
and does not depend on the external momentum. Hence, it is isolated  
in first term on the right-hand side of the equation of
Fig.~\ref{fig3}, while the difference
in the brackets is finite and can be computed in three dimensions. 
The first term factorizes into a three-loop diagram at 
${\bfm p}=0$, which has to be computed in $d$ dimensions, and 
the leading-order expression for $G(E)$. The divergent part 
cancels the contribution from $c_v$, and the remainder can 
again be evaluated in three dimensions. Thus, 
we do not need the $d-1$ dimensional Coulomb Green function, which 
is unknown.

\subsection{Result}

We write the perturbative expansion for the wave function at the
origin as 
\begin{eqnarray}
|\psi_n(0)|^2=|\psi^C_n(0)|^2\left(1+\delta^{(1)}\psi_n
+\delta^{(2)}\psi_n+\ldots\right)\,,
\label{series}
\end{eqnarray}
where the leading-order Coulomb wave function at the origin 
in three dimensions is 
given by $|\psi^C_n(0)|^2=(m \alpha_s C_F)^3/(8\pi n^3)$, and
$\delta^{(m)}\psi_n$ stands for the $m$th order correction. For the
ultrasoft part of the third-order correction we obtain
\begin{eqnarray}
\delta^{us}\psi_n&=&\,{\alpha_s^3\over\pi}\Bigg\{\!
\,\Bigg[{1\over 4}C_A^2C_F+{7\over 12}C_AC_F^2+{1\over
    6}C_F^3\Bigg]{1\over \epsilon^2}
+\Bigg[{1\over 6}C_A^2C_F+{1\over 2}C_AC_F^2+{1\over
    3}C_F^3\Bigg]{1\over \epsilon}\ln{\mu\over m}
\nonumber
\\
&&
\hspace*{-1cm} 
+\left[\left({5\over 18}-{\ln 2\over 6}\right)C_A^2C_F
+\left({25\over 12}-{5\over 6}\ln 2\right)C_AC_F^2+\left({19\over
    18}- \ln 2\right)C_F^3\right]{1\over \epsilon}
\nonumber\\
&&
\hspace*{-1cm} 
+\Bigg[-2C_A^2C_F-{16\over 3}C_AC_F^2-{8\over
    3}C_F^3\Bigg]\ln^2{\alpha_s}
+\Bigg[-\frac{5}{6}C_A^2C_F-{11\over 6}C_AC_F^2-{1\over
    3}C_F^3\Bigg]\ln^2{\mu\over m}
\nonumber\\
&&
\hspace*{-1cm} 
+\Bigg[{8\over 3}C_A^2C_F+{20\over 3}C_AC_F^2+{8\over
    3}C_F^3\Bigg]\ln{\alpha_s}\ln{\mu\over m}
\nonumber\\
&& 
\hspace*{-1cm} 
+\Bigg[C_A^3+\left({52\over 9}-{8\over 3}\ln2
-4H_{n}\right)C_A^2C_F
+\left(6
-{10\over 3n^2}
-{4\over 3}\ln2
-\frac{32}{3}H_{n}
\right)C_AC_F^2
\nonumber\\
&&
\hspace*{-0.5cm}+\left(-{52\over 9}
-{4\over 3n^2}
+8\ln2
-{16\over 3}H_{n}
\right)C_F^3\Bigg]\ln\alpha_s
\nonumber\\
&&
\hspace*{-1cm} 
 +\Bigg[-{3\over 4}C_A^3+\left(-{11\over 3}+{5\over 3}\ln2
+{8\over 3}H_{n}\right)C_A^2C_F
+\left(-{3\over 2}
+{5\over 3n^2}
+{1\over 3}\ln2
+{20\over 3}H_{n}
\right)C_AC_F^2
\nonumber\\
&&
\hspace*{-0.5cm} 
+\left(5
+{2\over 3n^2}
-6\ln2
+{8\over 3}H_{n}
\right)C_F^3\Bigg]\ln{\mu\over m}+\delta^{us}_n\Bigg\}\,,
\label{ultrasoft}
\end{eqnarray}
\begin{table}
  \begin{center}
    \begin{tabular}{|c|c|c|c|c|c|c|}
 \hline
 {$n$}& {1}& {2}& {3}& {4}& {5} & 6 \\
 \hline
 $\delta^{us}_n$ & 353.06 & 256.62 & 224.26 & 206.88 & 195.48 & 187.16\\
 \hline
    \end{tabular}
   \caption{\label{tab1} \small Numerical result
   for the non-logarithmic part of the ultrasoft contribution,
   as defined in (\ref{ultrasoft}); $n$ denotes the principal quantum 
   number.}
  \end{center}
\end{table}
\hspace*{-0.16cm}where 
$H_n=\ln\frac{C_F}{n}-\frac{1}{n}+\sum_{k=1}^{n-1}{1\over
  k}$. The most difficult part of the
calculation is the non-logarithmic term $\delta^{us}_n$, which 
we could compute only numerically. For the six lowest states its value
is given in Table~\ref{tab1}. Note that $|\psi^C_n(0)|^2$ in 
(\ref{series}) is formally defined in $d$ dimensions, but as explained
above, we  do not need the explicit $d$-dimensional expression, because 
the $1/\epsilon$ pole terms in (\ref{ultrasoft}) cancel with pole 
terms from $c_v^2 \,|\psi^C_n(0)|^2$ contained in~(\ref{vacpolnr}). 

We verified that the logarithmic terms in (\ref{ultrasoft}) 
when combined with those from the 
third-order potentials insertions~\cite{BKS2} agree 
with~\cite{KPSS2}\footnote{In Eq.~(7) of
  \cite{KPSS2} for the excited states the term
  $4C_F^3(1-1/n^2)/3$ is missed and the $\beta_0\ln{n}$ term should be 
  multiplied by three.}. 
The sum of the $1/\epsilon$ poles in the complete third-order correction 
to $|\psi_n(0)|^2$ from (\ref{ultrasoft}) and~\cite{BKS2} combined  
contains a term proportional to the single insertion of the 
first-order Coulomb potential and a term that must cancel 
against the infrared pole in twice the 
three-loop correction to $c_v$. This second term reads 
\begin{eqnarray}
\delta^{(3)}\psi_n^{\rm div}&=&{\alpha_s^3\over\pi}\,\Bigg\{
\left({1\over 36}C_A^2C_F+{5\over 48}C_AC_F^2+{5\over
    72}C_F^3\right)
\left({1\over \epsilon^2}+\frac{6}{\epsilon}\,\ln{\mu\over m}\right)
\nonumber\\
&&
\hspace*{-1cm} 
-\,\left({1\over 24}C_AC_F+{1\over 36}C_F^2\right){\beta_0\over
\epsilon^2} +\Bigg[\left({4\over 27}+{\ln2\over 2}\right)C_A^2C_F+
\left({113\over 162}+{\ln2\over 2}\right)C_AC_F^2
\nonumber\\
&& 
\hspace*{-1cm} 
+ \, \left({43\over
72}-\ln2\right)C_F^3 - {37\over 216}C_AC_FT_Fn_f+ 
{1\over 30}C_F^2T_F-{25\over
162}C_F^2T_Fn_f \Bigg]\,{1\over \epsilon} \,\Bigg\}\,,
\label{totalpol}
\end{eqnarray}
where $\beta_0=11C_A/3-4T_Fn_f/3$ is the one-loop QCD beta-function,
$T_F=1/2$ and $n_f$ is the number of light-quark flavors.
The $n_f$-part of the three-loop coefficient $c_v^{(3)}$ 
is known \cite{MPSS}, and we checked that the 
$n_f$-dependent pole parts cancel as required. 
Eq.~(\ref{totalpol}) is consistent with the scale dependence of the
hard matching coefficient given by Eqs.~(8), (9) of~\cite{KPSS2}, 
except for the rational part of the $C_AC_F^2$ term in
$\gamma_v^{\prime(3)}$, which should be increased by $7/24$ to agree
with our result.\footnote{In addition, as noted in~\cite{MPSS}, 
the term $-\frac{3}{2}\beta_0\gamma_v^{(2)}$ in Eq.~(8) 
of~\cite{KPSS2} should read  $+\beta_0\gamma_v^{(2)}$.}
 The difference is due to the fact that by using 
$[\sigma^i,\sigma^j]=i \epsilon^{ijk}\sigma^k$ the spin-algebra 
in~\cite{KPSS2} is not completely $d$-dimensional, hence 
the result for the scale-dependence of $c_v^{(3)}$ given there does not 
correspond to a calculation in conventional dimensional regularization.

\section{Quarkonium phenomenology}
\label{sec3}

We briefly discuss the size of the ultrasoft correction for the two 
most relevant cases, the leptonic decay of the $\Upsilon(1S)$ 
and the threshold production of top quark-antiquark pairs in 
$e^+e^-$ annihilation. This discussion must necessarily be 
preliminary, since the ultrasoft correction alone is factorization 
scheme and scale dependent. For the quarkonium spin-triplet 
ground state, $n=1$, we obtain from (\ref{ultrasoft}), omitting 
the $1/\epsilon$ poles,  
\begin{eqnarray}
\delta^{us}\psi_1 &=& \alpha_s^3\,\bigg\{
-18.71 \ln^2\alpha_s+52.03 \ln\alpha_s+112.38 
\nonumber\\
&& \hspace*{0.9cm} 
+\, \Big[23.52\ln\alpha_s-30.98\Big]\,\ln\frac{\mu}{m} 
-6.55 \ln^2\frac{\mu}{m}\bigg\}. 
\end{eqnarray}
The scale of the coupling $\alpha_s$ is 
most naturally of order of the inverse Bohr radius $m \alpha_s C_F$ 
in two of the three powers of the overall factor $\alpha_s^3$, 
and of order of the 
ultrasoft scale $m \alpha_s^2$ in the third. However, any other scale choice 
is formally equivalent at this order. In the following we evaluate 
$\alpha_s$ at $\mu_B=mC_F\alpha_s(\mu_B)$, wherever it appears. The scale
$\mu$ in the $\ln\,(\mu/m)$ terms 
is related to scale-dependent potentials and hard matching 
coefficients. We vary $\mu/m$ between 
$\alpha_s C_F$ (corresponding to the scale $\mu_B$) and 
1 (hard scale).  

\paragraph{\it $\Upsilon(1S)$ leptonic width.} Up to a
normalization factor the $\Upsilon(1S)$ leptonic decay width is given by
the residue of $\Pi(q^2)$ at the ground state
pole.  The leading-order expression for the decay width follows from
(\ref{vacpolnr}), resulting in 
$\Gamma_1^{\rm LO}=4\pi N_c e_b^2\alpha^2|\psi_1^C(0)|^2/
\left(3m_b^2\right)$, where $e_q$ is the electric charge of 
the quark flavour $q$
and $\alpha$ is the fine-structure constant.  The
non-perturbative contribution to the width is quite sizeable and out of
control for higher resonances~\cite{TitYnd,Pin}, hence we consider 
only the case $n=1$. Adopting $\alpha_s=0.30$, which corresponds to 
$\mu_B\approx 2\,$GeV for the $\Upsilon(1S)$, we
obtain $\delta\Gamma_1/\Gamma_1\approx 3.0$ from the 
non-logarithmic correction $\delta_1^{us}$ alone. Proceeding as
described above for the logarithmic terms results in the estimate 
\begin{equation}
  \delta\Gamma_1 \approx \Big[0.61-1.93\Big]\,\Gamma_1^{\rm LO}.
  \label{gamnum}
\end{equation}
It therefore appears that the large non-logarithmic term leads 
to a large enhancement of the width. Whether or not perturbation 
theory is out of control (as may be suggested by the upper limit of
the given range) can be decided only after combining all
third-order terms. 

\paragraph{\it  Top-quark production near threshold.} For 
top quarks non-perturbative effects are negligible, but its 
decay width $\Gamma_t$ smears out the Coulomb resonances below 
the threshold.  The NNLO analysis of the
cross section \cite{gang} shows that only the ground-state pole gives
rise to a prominent resonance. Although the calculation of the 
normalized cross section $R=\sigma(e^+e^-\to t\bar
t X)/\sigma(e^+e^-\to\mu^+\mu^-)$ requires 
the full Green function $G(E)$, the height of the resonance 
can be estimated from the wave function at the origin of the 
{\it would-be} toponium ground state. In the leading-order 
approximation $R_1^{\rm LO}\approx 6\pi
N_ce_t^2|\psi_1^C(0)|^2/\left(m_t^2 \Gamma_t\right)$.  Adopting 
$\alpha_s=0.14$, which corresponds to $\mu_B\approx 32.5\,$GeV, we
obtain $\delta R_1/R_1\approx 0.31$ from the 
non-logarithmic correction $\delta_1^{us}$ alone. 
Including an estimate of the logarithmic terms we find
\begin{equation}
  \delta R_1 \approx \Big[(-0.17)-(+0.13)\Big]\,R_1^{\rm LO}.
  \label{rnum}
\end{equation}
Hence, despite the large quark mass, we may anticipate a sizeable 
third-order correction, unless there are cancellations. 

\section{Summary}
\label{sec4} 
We evaluated the third-order correction to electromagnetic 
quarkonium production and annihilation due to the emission and 
absorption of an ultrasoft gluon. Together 
with other contributions~\cite{PSS,BKS1,KniPen2,ManSte,KPSS2,Hoa2,BKS2,MPSS} 
already completed 
the problem of evaluating the total ${\cal O}(\alpha_s^3)$ corrections
is now reduced to the calculation of four ${\cal O(\epsilon)}$ 
terms in the NNNLO heavy quark-antiquark potential \cite{BKS2}, 
the three-loop colour-singlet Coulomb potential, 
and the three-loop vector current
matching coefficient in the $\overline{\rm MS}$ scheme in 
dimensional regularization. The previously
unknown non-logarithmic ultrasoft contribution is large and significantly
increases the production and annihilation rates. It might limit the 
accuracy of the perturbative analysis of the quarkonium even for 
top quarks.  We should however emphasize that a definite conclusion 
can only be drawn once the full NNNLO result is available. 
In this respect the sizable negative third-order 
correction from the perturbation potentials \cite{PSS,BKS1,BKS2} 
should be mentioned.

\vspace*{1em}

\noindent
\subsubsection*{Acknowledgement}
This work is supported by the DFG Sonder\-forschungsbereich/Transregio~9 
``Computergest\"utzte Theoretische Teilchenphysik''. 
The work of A.A.P. is supported  by BMBF Grant No.\ 05HT6VKA.

\end{document}